# Decision Trees for Intuitive Intraday Trading Strategies

Naga Prajwal
UG Student Department of CSE
The National Institute of Engineering
Mysuru, India
nagaprajwalnpb@gmail.com

Dinesh Balivada
UG Student Department of CSE
The National Institute of Engineering
Mysuru, India
dineshbalivada8@gmail.com

Sharath Chandra Nirmala
UG Student Department of CSE
The National Institute of Engineering
Mysuru, India
sharathnimala@gmail.com

Tiruveedi Poornoday
UG Student Department of CSE
The National Institute of Engineering
Mysuru, India
tpoornoday01@gmail.com

*Abstract*—**This research paper aims to investigate the efficacy of decision trees in constructing intraday trading strategies using existing technical indicators for individual equities in the NIFTY50 index. Unlike conventional methods that rely on a fixed set of rules based on combinations of technical indicators developed by a human trader through their analysis, the proposed approach leverages decision trees to create unique trading rules for each stock, potentially enhancing trading performance and saving time. By extensively backtesting the strategy for each stock, a trader can determine whether to employ the rules generated by the decision tree for that specific stock. While this method does not guarantee success for every stock, decision tree-based strategies outperform the simple buy-and-hold strategy for many stocks. The results highlight the proficiency of decision trees as a valuable tool for enhancing intraday trading performance on a stock-by-stock basis and could be of interest to traders seeking to improve their trading strategies.**

## I. Introduction

Intraday trading involves buying and selling stocks within the same day to benefit from small price movements in the market, yielding small profits that add up over the trading period. Technical analysis is a long-established method in intraday trading that utilizes past market data to create indicators, identify patterns, and make trading decisions based on the observed patterns. However, traditional technical analysis methods rely on a fixed set of rules based on combinations of technical indicators, which tend to be tediously generated and may not perform well for all stocks. Furthermore, these methods may not consider individual stock characteristics, which can result in suboptimal trading decisions.

Decision Trees can serve as an alternative approach to the manual construction of trading rules in a trading strategy. They create unique and interpretable trading rules for each stock, potentially enhancing trading performance and saving time. Additionally, decision trees can perform classification and regression tasks [1].

This paper demonstrates the efficacy of decision trees as a valuable tool for enhancing intraday trading performance on a stock-by-stock basis. Despite using the same indicators for each stock, a decision tree-based classifier model discovers different rules depending on the stocks' characteristics. Furthermore, decision trees can create interpretable and understandable models, helping traders better understand and interpret their trading strategies.

## II. Dataset Collection

The dataset collection methodology in this paper adopts a hybrid approach, which efficiently and economically acquires comprehensive intraday trading data. The approach utilizes two vendors, namely ICICI Breeze and Yahoo Finance. The ICICI Breeze API was the primary source for data, providing one-minute time interval data from 2022 to 2024. However, ICICI Breeze data lacks adjusted close prices, which is essential for backtesting. The Yahoo Finance API is used to surpass this limitation. Yahoo Finance's daily interval data, which includes adjusted close prices, is used to compute adjustment factors for each trading day. These adjustment factors are resampled to a one-minute frequency, aligning with the one-minute interval of the ICICI Breeze data. This procedure ensures the synchronization of adjusted close prices with the ICICI Breeze data, facilitating a thorough analysis of intraday price movements. The hybrid approach used in this paper enables the acquisition of adjusted one-minute data over the past two years at no cost, making it a quick, cost-effective method for retail traders. By utilizing this approach, traders can access comprehensive historical intraday trading data, which aids in making informed trading decisions.

## III. Strategy Building

### A. Indicators Used

The strategy-building process involved four traditional Technical Indicators, and five statistical relationships between the prices, resulting in nine inputs for the Decision Tree model. These indicators include the returns of the close price series, the 15-period return of the close price series, the 14-period Relative Strength Index (RSI) Indicator, the 14-period Average Directional Index (ADX), the ratio between the 14-period Simple Moving Average (SMA) and the close price series, the correlation coefficient between the SMA and the close prices, the 14-period rolling volatility, the 210-period rolling volatility (which is essentially the 14-period volatility of the 15-period rolling returns), and the ratio between the 14 period rolling Volume Weighted Average Price (VWAP) and the Close Price Series.

The indicators were chosen as they provide meaningful information about the stocks' current and probable future performance. The returns of the close price series and the 15-period return of the close price series were included to capture the price changes over short and medium-term horizons. The 14-period RSI and ADX were used to assess the momentum and direction of the trend, respectively. The ratio between the 14-period SMA and the close price series and the correlation coefficient between the SMA and the close prices were included to capture the long-term price trend. The 14-period rolling volatility and the 210-period rolling volatility were used to capture the market volatility over different periods. Finally, the ratio between the 14-period rolling VWAP and the close price series was included to assess the volume-weighted average price and its relationship with the price change.

Overall, the chosen indicators provide a comprehensive and diverse collection of inputs for the Decision tree-based classifier model, enabling the creation of unique and interpretable trading rules for each stock.

*B. Decision Trees*

Decision trees, a widely utilized supervised machine learning model, find extensive application across diverse domains such as finance, medicine, and marketing. This model operates by recursively partitioning the dataset into subsets based on the input variables. Within the tree structure, nodes represent the input variables, branches depict the potential values of these input variables, and leaves correspond to the output variables [2].

Classification trees are a specific type of decision tree, frequently employed in financial models to generate discrete buy and sell signals. This model produces a binary output, where a value of 1 indicates a buy signal and a value of 0 indicates a sell signal. To create the output for the model, historical data of the stock's closing price returns, shifted by -1, are utilized. In this context, future negative returns map to 0 and future positive returns to 1 [3].

*1) Criteria for Building Decision Trees*
Criteria for building decision trees can be entropy, Gini coefficient, or classification error. These criteria measure node impurity, which indicates the extent to which output variables are mixed within a node. The criterion used to build the decision tree determines the splitting rule used to partition the data at each node.

*2) Gini Coefficient as the Preferred Criterion*
This paper employs the Gini coefficient for constructing decision trees. The Gini coefficient assesses the probability of misclassifying a randomly selected sample from a node, based on the distribution of the output variables within that node. It was chosen over other criteria due to its lower sensitivity to the distribution of output variables and its capacity to manage imbalanced datasets effectively. Moreover, the Gini coefficient typically generates smaller trees, thereby mitigating the risk of overfitting the model.

Optimizing the depth of the decision tree is essential, as it is a critical hyperparameter that affects both the model's complexity and performance. A tree depth that is too shallow can render the model overly simplistic, preventing it from capturing the underlying patterns in the data. This can lead to underfitting, where the model performs inadequately on both training and testing data. In such cases, the decision boundaries may be overly simplified, diminishing the model's predictive accuracy. Conversely, an excessively deep tree can make the model overly complex and susceptible to overfitting. Overfitting occurs when the model memorizes the training data instead of identifying the underlying relationships, resulting in excellent performance on training data but poor generalization to new data.

For our paper, we conducted experiments with decision tree depths ranging from three to six. After thorough analysis, we determined that a depth of four provides the optimal balance for model complexity and performance. Charts of performance metrics concerning tree depth are provided below from Fig. 1 to Fig. 3. A tree depth of four enables the model to capture sufficient complexity for accurate predictions while minimizing the risk of overfitting.

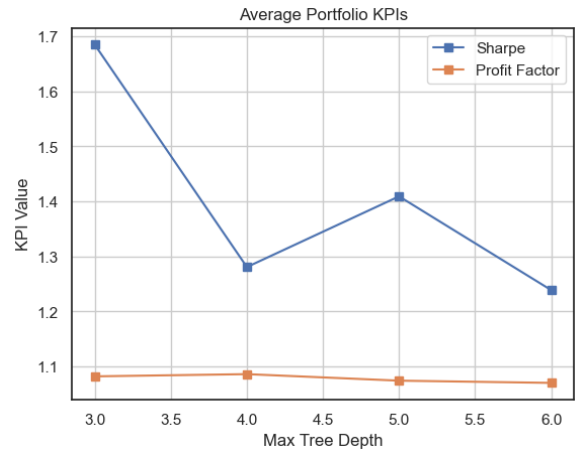

Fig. 1  Sharpe and Profit Factor of an average portfolio w.r.t tree depth (testing dataset).

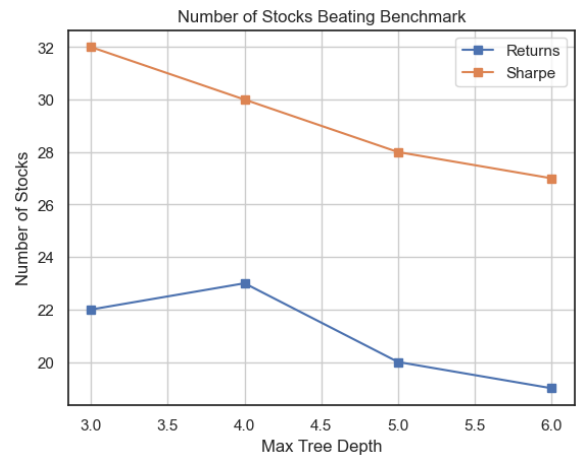

Fig. 2  Number of stocks where strategy outperformed benchmark (testing dataset).

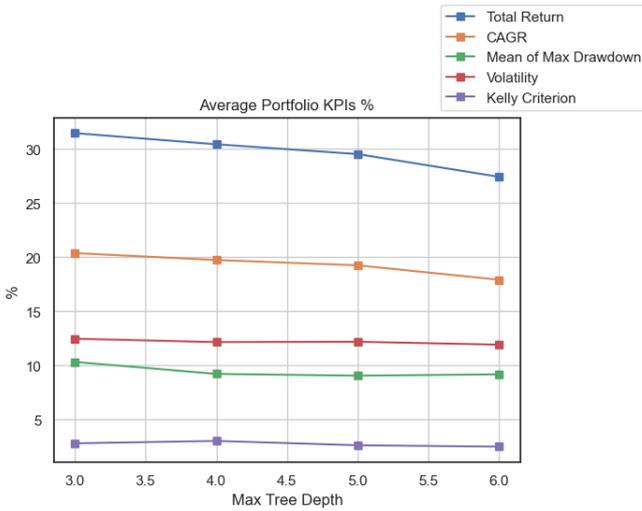

Fig. 3     Additional KPIs w.r.t tree depth (testing dataset)

## IV. STRATEGY IMPLEMENTATION

### A. Dataset Preprocessing

The adjusted dataset constructed from various sources, as discussed in the "Dataset Collection", forms the basis for the intraday trading strategy. However, before proceeding with model training, it is necessary to preprocess the dataset to ensure consistency and continuity across all stocks.

All stocks are aligned to have an identical number of rows and indexes with matching values. Due to varying market conditions and brokers' data quality, certain stocks may exhibit missing values for a few timestamps. A combined index for all stocks was created by applying the set union operator to all stocks' indexes. The set union operator creates new indexes with empty values in some stocks, resulting in NaN (Not a Number) values. We fill these NaN values using linear interpolation to ensure data continuity for backtesting. This method is relatively safe as it maintains data integrity while filling relatively small gaps with interpolated values.

After aligning the data and interpolating the missing values, we apply the aforementioned technical indicators to the dataset. These indicators serve as inputs to the decision tree model, offering valuable insights into market trends and price movements. Additionally, the necessary output value for the model, indicating the trading signal (1 for buy and 0 for sell), is incorporated into the dataset. Since decision trees are not sensitive to monotonic transformations [6], there is no requirement to scale the features or transform the data. Consequently, the raw features are utilized directly in the model without preprocessing. Subsequently, the preprocessed data is divided into training and testing sets, with the training set spanning from 2022 to 2023, and the testing set spanning from 2023 to 2024. The training set is utilized to train the decision tree model, while the testing set is employed to evaluate its predictive performance on unseen data.

### B. Model Implementation

The decision tree-based intraday trading model is developed using the DecisionTreeClassifier module from the Scikit Learn library. This module provides an efficient and versatile implementation of decision trees for classification tasks. Through experimentation, a maximum depth of four is chosen for the decision tree, striking a balance between capturing adequate complexity for accurate predictions and avoiding overfitting. The Gini coefficient serves as the criterion for constructing the decision tree model, effectively measuring the impurity of a node's class distribution, which is particularly suitable for binary classification tasks. Despite providing nine inputs to the decision tree model, it is observed that, on average, most models utilize approximately five to seven indicators to make actual decisions. Moreover, these inputs exhibit variability across different stocks, indicating the model's adaptability in selecting relevant features for each stock. Fig 4 and Fig 5 depict the decision trees for the two stocks. An immediate distinction is the absence of the ADX as a deciding parameter in Fig. 4.

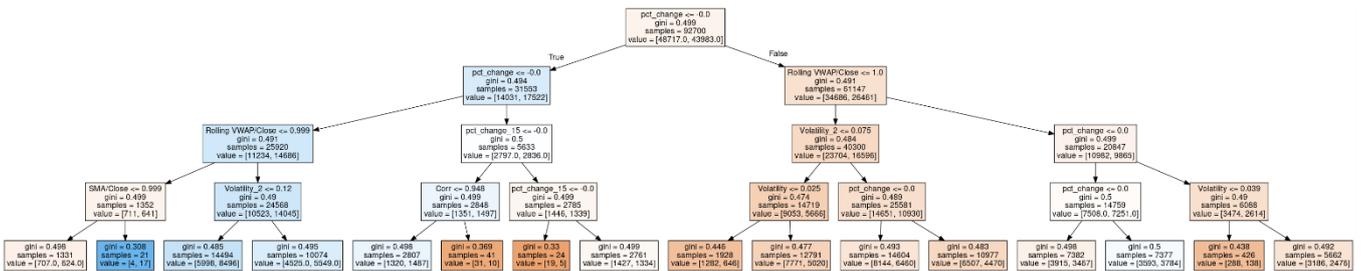

Fig. 4     Decision Tree Model for Nestle India's stock.

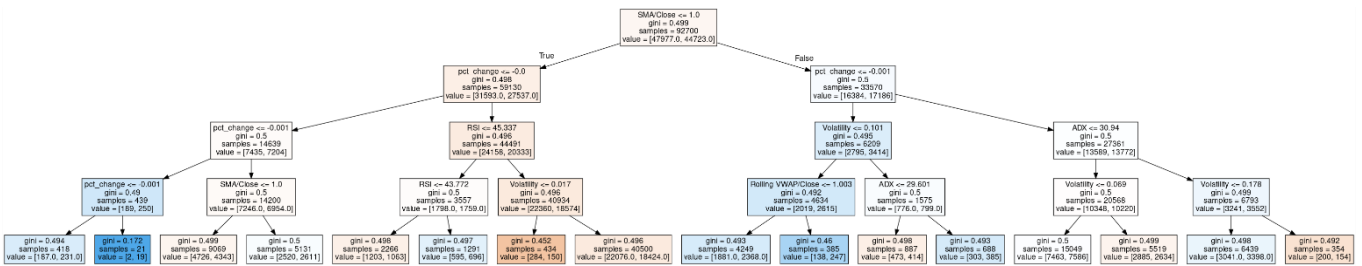

Fig. 5     Decision Tree Model for Reliance's Stock

*C. Backtesting*

Backtesting played a crucial role in evaluating the performance and robustness of the decision tree-based intraday trading strategy. The backtesting process and results were implemented using Python's vectorbt library for vectorized backtesting. The backtesting process involved simulating the execution of trading signals generated by the decision tree model on historical data. Python's vectorbt library provided efficient tools for vectorized backtesting, enabling fast and convenient evaluation of trading strategies. It's important to note that commissions and slippage were not explicitly incorporated into the backtesting process. However, the buying and selling signals were adjusted forward by one step to accommodate the trading delay, thereby simulating a realistic trading scenario. Returns were computed using close prices. To assess its relative performance, the decision tree-based intraday trading strategy was compared to a buy-and-hold benchmark. The outcomes of the backtesting offered insights into the profitability, risk-adjusted return, and consistency of the decision tree-based intraday trading strategy. Through the analysis of performance metrics, traders could make well-informed decisions regarding the potential implementation of the strategy in real-world trading scenarios. Overall, the backtesting process played a pivotal role in evaluating the effectiveness and efficacy of the decision tree-based intraday trading strategy, furnishing valuable insights for potential deployment in live trading environments.

## V. Results

Key performance indicators (KPIs) assume a pivotal role in the evaluation of trading strategy efficacy and performance. Serving as invaluable metrics, these KPIs facilitate the assessment of performance and risk characteristics inherent in the decision tree-based intraday trading strategy. The Python open-source libraries "vectorbt" and "quantstats" furnish these KPIs for individual stocks. It is pertinent to mention that the annualizing of KPIs utilizes $252 \times 375 = 94500$ periods per year [4], with 252 representing trading days per year and 375 denoting trading intervals within a trading day. The risk-free rate is assumed to average 7.2% [5]. The ensuing section delineates the KPIs employed in the analysis:

1) *Sharpe Ratio*
   Measures the risk-adjusted return of an investment, indicating how much return an investment generates per unit of risk.

2) *Total Return*
   Represents the overall return generated by an investment over a specified period, including capital appreciation and income.

3) *CAGR (Compound Annual Growth Rate)*
   Represents the geometric average annual rate of return over a specified period, providing a smooth annualized growth rate.

4) *Maximum Drawdown*
   Measures the maximum loss from peak to trough experienced by an investment or trading strategy over a specific period.

5) *Win Rate*
   Represents the total profitable trades relative to the total trades executed, indicating the consistency of a trading strategy's performance.

6) *Profit Factor*
   Measures the relationship between profits and losses, calculated by dividing the total profit from winning trades by the total loss from losing trades.

7) *Volatility*
   Measures the degree of variation or dispersion in the returns of an investment or trading strategy over time, indicating the riskiness or stability of returns.

8) *PSBBR*
   The percentage of Stocks beating their benchmark (buy-and-hold) in total returns metric.

9) *PSBBS*
   The percentage of Stocks beating their benchmark (buy-and-hold) in Sharpe Ratio metric.

The strategy demonstrated promising results, outperforming the benchmark in terms of Sharpe ratio, volatility, and maximum drawdown. However, it's noteworthy that the total returns of the strategy were lower than the benchmark in the testing dataset. Despite this, when adjusted for volatility, the strategy outperformed the benchmark, delivering a better reward-to-risk ratio and lower maximum drawdown.

These findings illustrate that despite the strategy's total returns being marginally lower in the testing dataset, its risk-adjusted performance is superior to the benchmark. This indicates that the decision tree-based intraday trading strategy is more efficient in terms of managing risk and delivering consistent returns. This is further highlighted by PSBBR and PSBBS values drop from the training to the testing dataset, but the drop is far more significant for PSBBR at a 42.50% drop compared to a drop of 28.57% for PSBBS. The results for both the training and testing data have been tabulated below in Table 1. Additionally, Fig. 6 to Fig. 11 are charts to aid visualization of the returns.

TABLE 1    Final Results

| KPIs Table | Training Dataset | | | Testing Dataset | | |
|---|---|---|---|---|---|---|
| | *Benchmark* | *Strategy* | *Under/Overperforms* | *Benchmark* | *Strategy* | *Under/Overperforms* |
| **Sharpe Ratio** | 0.1 | 3.37 | OVER | 2.36 | 5.81 | OVER |
| **Total Return [%]** | 7.17 | 26.9 | OVER | 34.23 | 28.62 | UNDER |
| **CAGR [%]** | 4.96 | 18.15 | OVER | 22.95 | 19.33 | UNDER |
| **Max Drawdown [%]** | -14.31 | -2.33 | OVER | -7.02 | -2.63 | OVER |
| **Win Rate [%]** | NA | 48.20 | NA | NA | 48.15 | NA |
| **Profit Factor** | 1.08 | 1.07 | UNDER | 1.63 | 1.11 | UNDER |
| **Volatility [%]** | 18.27 | 5.21 | OVER | 10.23 | 3.32 | OVER |
| **PSBBR [%]** | NA | 80 | NA | NA | 46 | NA |
| **PSBBS [%]** | NA | 84 | NA | NA | 60 | NA |

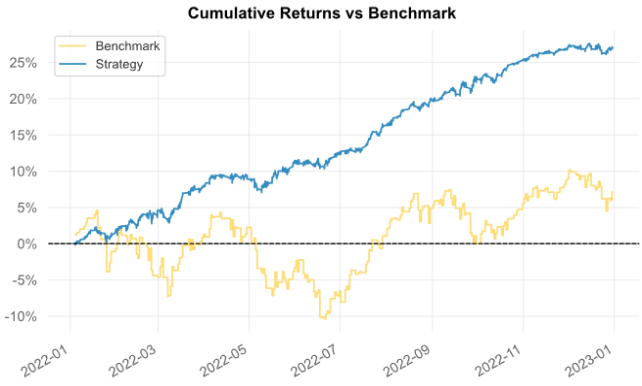

Fig. 6    Returns of Strategy and. Benchmark (Training Dataset)

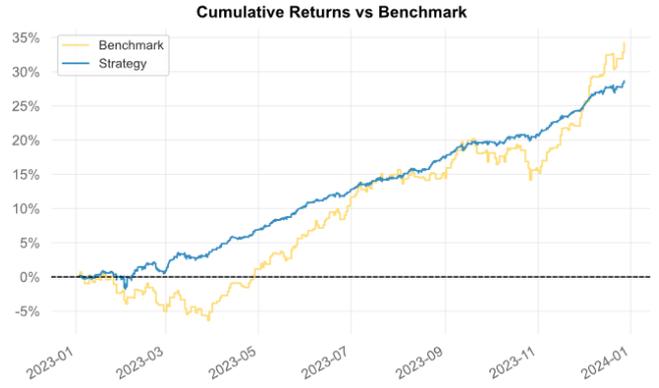

Fig. 9    Returns of Strategy and. Benchmark (Testing Dataset)

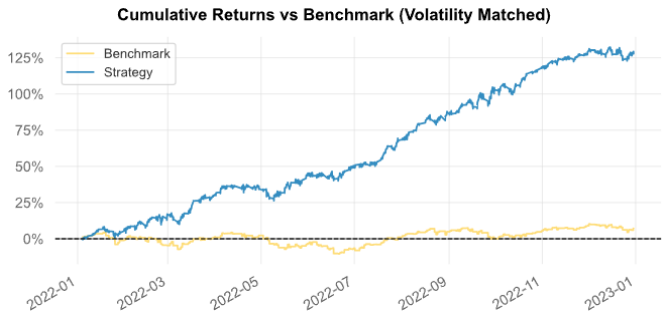

Fig. 7    Volatility adjusted returns of Strategy and Benchmark (Training Dataset)

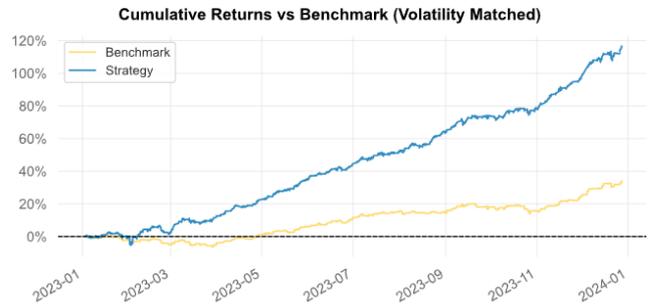

Fig. 10    Volatility adjusted returns of Strategy and Benchmark (Testing Dataset)

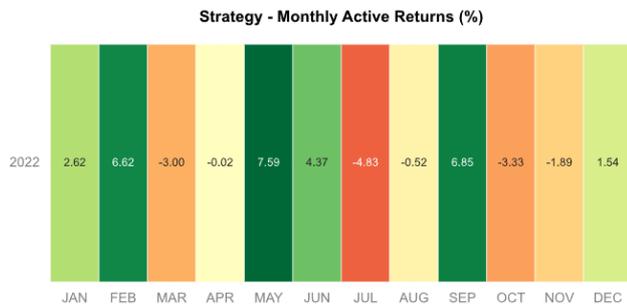

Fig. 8    Monthly Returns of Strategy (Training Dataset)

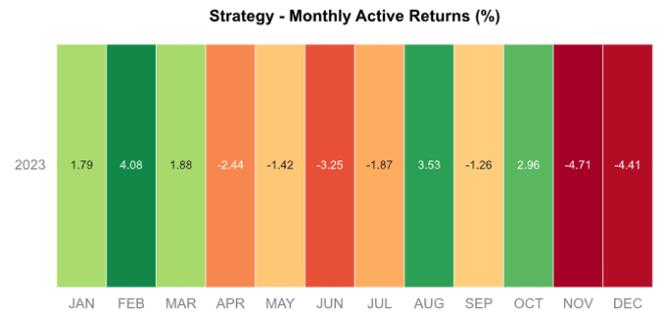

Fig. 11    Monthly Returns of Strategy (Testing Dataset)

## VI. CONCLUSION

This paper has demonstrated the effectiveness of decision trees in creating intraday trading strategies with technical indicators. The study found that decision trees offer a powerful and efficient method for generating trading rules based on a combination of technical indicators, providing traders with valuable insights and actionable signals.

The research provides evidence that decision trees can efficiently select relevant features from a pool of technical indicators, making them adaptable to different stocks and market conditions. Moreover, decision trees-based classifier outperformed the buy-and-hold benchmark in terms of efficiency and adaptability. By leveraging decision trees and technical indicators, traders can make better trading decisions in dynamic market environments.

Future enhancements to the strategy could focus on portfolio-level measures to boost overall performance. For instance, pruning stocks from the universe that don't outperform their benchmark could enhance overall portfolio performance. In our case, removing five underperforming stocks improved the strategy's performance. Such pruning would be particularly essential for longer periods and larger universes of stocks.

Furthermore, while this paper tested only a few indicators, there are hundreds of indicators available to create new trading strategies. Future research could explore the effectiveness of different combinations of indicators and refine the decision tree-based approach for optimal performance.

## ACKNOWLEDGEMENTS

We would like to express our sincere gratitude to the following mentors and professors for their invaluable guidance and support throughout this research:

- Amruthasree V M, Assistant Professor, Department of CSE, The National Institute of Engineering
- Mahe Mubeen Akhtar D, Assistant Professor, Department of CSE, The National Institute of Engineering
- Ramesh G, Assistant Professor, Department of CSE, The National Institute of Engineering

Their expertise and encouragement have been instrumental in shaping this research work. We are grateful for their mentorship and unwavering support.